\documentstyle[prb,aps,epsfig]{revtex}
\def\E{{\rm e}}
\def\I{{\rm i}}
\def\D{{\rm d}}
\def\BI{{\rm I}}
\begin{document}

\title{Effect of a forbidden site on a $d$-dimensional lattice random walk}
\author{Nicolas Martzel \footnote
  {{\bf e-mail:} martzel@gps.jussieu.fr} and Claude Aslangul\footnote
  {{\bf e-mail:} aslangul@gps.jussieu.fr}}
\address{Groupe de Physique des Solides, Laboratoire associ\'e au
CNRS UMR 7588, \\ Universit\'es Paris 7 \&
Paris 6, Tour 23, Place Jussieu, 75251 Paris Cedex 05, France}

\date{\today}

\maketitle
\begin{abstract}
We study the effect of a single excluded site on the diffusion
of a particle undergoing random walk in a $d$-dimensional lattice.
The determination of the characteristic function
allows to find explicitly the asymptotical behaviour of physical quantities
such as the particle average position (drift) $\langle\,\vec x\rangle(t)$
and the mean square deviation
$\langle\,\vec x^{\,2}\rangle(t)-\langle\,\vec x\rangle^2(t)$.
Contrarily to the one-dimensional case, where $\langle\,\vec x\rangle(t)$
diverges at infinite times ($\langle\,\vec x\rangle(t)\,\sim
t^{1/2})$ and where the diffusion constant $D$ is changed due to the
impurity, the effects of the latter are shown to be much less
important in higher dimensions: for $d\,\ge\,2$, $\langle\,\vec x\rangle(t)$
is simply shifted by a constant and
the diffusion constant remains unaltered although dynamical
corrections (logarithmic for $d\,=\,2$) still occur.
Finally, the continuum space
version of the model is analyzed; it is shown that $d\,=\,1$, is
the lower dimensionality above which all the effects of the
forbidden site are irrelevant.
\end{abstract}


\section{Introduction}
\label{intro}

Interactions between diffusing particles are expected to play an
important role in many systems of physical interest (zeolites
\cite{kukla}, biological membranes \cite{nener,sack}, one-dimensional hopping
conductivity \cite{rich}), especially in low dimension and even with
short-range interactions. For $d=1$, the case of a contact
interaction has been extensively studied in the past, following the
pioneering paper by Harris \cite{harris} on the so-called tracer
problem, first showing that the mean-square displacement
$\langle x^{2}\rangle$ increases as $t^{1/2}$ at large times. This result was
also obtained by van Beijeren {\it et al.}\cite{vanBeij} and, more
recently, by R\"{o}denbeck{\it et al.}\cite{Rod98}; these authors
indeed solve the full $N$-particle problem for an arbitrary initial
condition and recover Harris' result by going to the
$N\rightarrow+\infty$ limit. Still in the case of a zero-range
repulsive interaction, asymptotic
results (transport coefficients and distribution
laws) in the extreme case of a finite-$N$ compact initial cluster
have been obtained in ref\cite{aslan}, where the relation with the
theory of extreme events was also discussed\cite{Gumb58}.

Obviously, the effects of short-range interactions are expected to become less
and less important in higher dimensions (on the other hand, the
interplay between interaction range and space
dimensionality remains, as far as we know, an open question). In
the present paper, we address the simplest problem, namely that of
two particles with a contact interaction undergoing $d$-dimensional lattice
random walk. After transforming to the center-of-mass frame, this
essentially  reduces to the Brownian motion of a single particle
subjected
to an isolated reflecting barrier. The equations for the general
$d$-dimensional case are given below and, among other things, allow to show
that $d=1$ is indeed the marginal dimensionality above which the
random walk is unaffected by the presence of the localized impurity in the
continuous version of the model; by this, it is meant that transport
coefficients,
as deduced from the dominant term in the
asymptotic regime of physical quantities, are the same with and without
impurity for $d\,>\,1$. Nevertheless, corrections still occur in the lattice
model; that these corrections, less and less
relevant as $d$ increases, basically display a $a^d$ behaviour
for $d\,>\,1$, where $a$ denotes the lattice spacing.

This paper is organized as follows. We first write down the basic
equations for any dimension $d$ and give the expression of the
Laplace transform of the generating function giving by derivations
all the moments of the particle coordinate in the lattice. We then analyze
this general result, retrieving well-known results for the $d=1$ case.
Subsequently, we examine in details the two-dimensional case, which exhibits
lattice-dependent logarithmic corrections, and give the main
results for the $d>2$ case. Even in such high dimensions, the
presence of the forbidden site has consequences on the asymptotical
dynamics: the final average particle position displays a shift $\delta
x_{\rm f}$ which scales as
$a^{d}x_{0}^{-d+1}$ where $x_{0}$ is the distance between the
starting point of the
particle and the exclusion site. Finally, the space-continuous version
solution is written in full, showing that the moment-generating
function is, in this limit, insensitive to the presence of the
excluded site as far as the dimensionality $d$ is strictly greater that
one.


\section{Basic equations for the {\it d}-dimensional lattice walk}
\label{prob}

In the following, we consider a particle undergoing random walk in
the viscous limit on a $d$-dimensional hypercubic lattice having a
forbidden site. Denoting $\vec e _{p}$
the unit-vector in the $p$-th direction of the lattice, an
arbitrary site of the lattice is fully specified by its vector $\vec
n\,=\,\sum_{p=1}^{d}
n_{p}\vec e _{p}$; the excluded site is located at the origin. Let
$P(\vec n ,t)$ be the probability to find the particule at site
$\vec n $ at time $t$. In the continuous time description, it is
readily seen that the time
evolution of $P(\vec n ,t)$ is governed by the following master equation:
\begin{equation}
     \frac{\D}{\D t}\,P(\vec{n},t)\,=\, W[1-\delta(\vec n,\vec
     0)]\,\sum_{\varepsilon=\pm 1} \sum_{p=1}^{d}\,
     \,[1-\delta(\vec n+\varepsilon\vec e_{p},\vec 0)]\,
     [P(\vec n+\varepsilon\vec e_{p},t)
     -\,P(\vec n,t)]\enspace, \label{masterd}
\end{equation}
where $W$ denotes the probability per
unit time to jump from one site to any of its nearest-neighbours; in
(\ref{masterd}),
$\delta(\vec n,\vec m)\,$ equals $1$ if $\vec n=\vec m$ and vanishes otherwise;
obviously, ${\dot P}(\vec n=0,t)=0$ at all times.
The easiest way to solve this problem is to deal with the characteristic
function of the probability distribution:
\begin{equation}
     \psi(\vec \phi,t)\,=\, \sum_{\vec n} \E ^{\I \vec \phi. \vec n} \,
      P(\vec n,t) \enspace, \label{characteristic}
\end{equation}
which allows to find the probability function $P$ by the inverse formula :
\begin{equation}
      P(\vec n,t)\,=\,\int\, \D^d \phi
      \,\E^{- \I \vec \phi.\vec n}  \psi(\vec\phi,t) \enspace,
     \label{defcharac}
\end{equation}
where $\phi_{p}=\vec \phi.\vec e_{p}$ and where $\int\, \D^d \phi$ means
$(2 \pi)^{-d} \int _{- \pi }^{+ \pi}\D
\phi_{1}\int _{-
\pi }^{+ \pi}\D \phi_{2}\dots\int _{- \pi }^{+ \pi}\D \phi_{d}$.
Multiplying both sides of (\ref {masterd}) by $ \E ^{\I \vec
\phi.\vec n}$ and summing over $\vec{n}$, one readily obtains:
\begin {equation}
     \frac{\partial}{\partial t}\,\psi(\vec{\phi},t) \,=\, 2 W \sum_{p=1}^{d}
     (\cos{\phi_{p}} - 1)
     \psi(\vec \phi,t) \,+\, W \sum_{\varepsilon=\pm
1}\sum_{\,p=1}^{d}\,(\E ^{\I \varepsilon\phi_{p}}-1) \,
      \int\, \D^d \phi'\,\left(\E^{- \I\varepsilon \phi'_{p}}-1\right)
\psi(\vec\phi\,',t) \enspace.
     \label{masterpsi}
\end{equation}
For sake of simplicity, the initial position of the walker is chosen to be
$\vec x_{0} = n_{0}a \,\vec e_{1}$. With this condition, and
since there is no external bias, the probability distribution is invariant
under a mirror transformation through any plane perpendicular
to $ \vec e_{p}$
with $p \neq 1$. In addition, the quantities $\int\, \D^d \phi'
 \,\cos { \phi'_{p}} \psi(\vec{\phi\,'},t)$ , $p \neq 1$, are all
equal. Introducing the Laplace transform of $\psi(\vec\phi,t)$ :
\begin {equation}
     \tilde{\psi}(\vec\phi,z)\,=\,\int_{0}^{+\infty}\,\D t\,\E ^{-zt}
     \psi(\vec\phi,t)
\enspace,
\label{laplace}
\end {equation}
one readily obtains:
\begin {eqnarray}
     \left(z + 2W \sum_{p=1}^{d}(1- \cos
     \phi_{p})\right)\tilde{\psi}(\vec{\phi},z)
     \,&=&\,\psi_{0}+2W\sum_{p=2}^{d}\,(\cos\phi_{p} - 1)
      \int\, \D^d \phi\,'
      \,(\cos \phi'_{p} -1)\tilde{\psi}(\vec{\phi\,'},z) \nonumber\\
      &+& \,2W\int \D^d \phi\,'
     \,[(\cos\phi_{1} - 1)\, \cos \phi'_{1} + \sin \phi_{1}\sin
     \phi'_{1}]\,\tilde{\psi}(\vec\phi\,',z)
\enspace,
\label{masterlap}
\end{eqnarray}
where $\psi_{0} \,\equiv\, \psi (\vec\phi,\,0) \,=\, \E ^{\I
 n_{0}\phi_{1}}$. Equation (\ref{masterlap}) is a homogeneous separable
Fredholm system, and
can be solved by quadratures (see Appendix
\ref{AppendixA}), yielding the expression of $\tilde {\psi} (\vec{\phi},z)$
(see eq. (\ref{solcharac}).
For further reference, we explicitly write down the Laplace transforms of
the average coordinate and mean-square displacement:
\begin{equation}
     \tilde{\langle x_{p} \rangle} (z) = - \I a \left. \frac { \partial
{\tilde{\psi} (\vec \phi,\,z)} }
      {\partial {\phi_{p}}}\right|_{\,\vec \phi = \vec 0 } \enspace,
\hspace{30pt}
      \tilde{\langle x_{p}^{2} \rangle} (z) = - a^2 \left. \frac { \partial^{2}
     {\tilde {\psi} (\vec \phi,\,z)}}
     {\partial \phi_{p}^{2}} \right|_{\,\vec \phi = \vec 0 }
\enspace.\label{genmom}
\end{equation}
With the solution given by (\ref{solcharac}), one precisely has:
\begin{equation}
     \tilde{\langle x_{p} \rangle} (z) = \delta_{p1}\,\frac{a}{z}
     \left[n_{0} - \I \tilde{S_{1}}(Z) \right]\enspace,
     \label{x1xp}
\end{equation}
\begin{equation}
     \tilde{\langle \vec x^{\,2} \rangle} (z) = \frac{a^{2}}{z}
     \left[ \frac{d}{ Z} + n_{0}^{2} + \tilde{C_{1}}(Z) +
     (d-1) \tilde{C}(Z) \right] \enspace,
\label{x2d}
\end{equation}
where $Z=z/(2W)$ and where the functions
$\tilde{S_{1}}$, $\tilde{C_{1}}$ and  $\tilde{C}$ are defined in Appendix
\ref{AppendixA}. In the next sections, we shall find the large-time
behaviour of these
quantities according to the dimensionality $d$, by analyzing their
small $Z$ expansions ($Z\,\ll\,1\,\,\Longleftrightarrow\,\,t\,\gg\,W^{-1}$).



\section{The one-dimensional case}
\label{oned}

The reduced problem is here simply the $1\,d$ diffusion
of a particle submitted to a reflecting barrier. Although the result
is well known, we shall discuss it to stress the main difference between
this case and that of larger dimensionality. For $d=1$ the relevant ${\tilde
C}_{1}$ function (\ref{fredsolC1S1}) greatly simplifies (${\tilde
C}_{1}\,=\,\beta_{n_{0}}/(1-a_{1}-\beta_{1})$),
whereas $\tilde{S}_{1}$ is formally unchanged and still given by
(\ref{fredsolC1S1}).
Let first us discuss the drift term, given by (\ref{x1xp}). By
using\cite{gradry}:
\begin{equation}
     \int_{0}^{\infty} \D x \,\, \E^{-\alpha x} \BI_{n} (x) =
     \frac{1}{(\alpha + \sqrt{\alpha^2 - 1})^n \sqrt{\alpha^2 - 1}}\enspace,
     \hspace {30pt}
     \int_{0}^{\infty} \,\D x\,x^{-1} \, \E^{-\alpha x} \BI_{n} (x) =
     \frac{(\alpha - \sqrt{\alpha^2 -1})^n}{n}\enspace,
     \label{defint1}
\end{equation}
we find that $\gamma_{n}$ (introduced in (\ref{fredbessela0b0}))
is well-defined  for all $z,\,\Re z\ge 0$ and
indeed has a finite limiting value, namely $\gamma_{n}(z=0)=1$;
this is one peculiarity of the one-dimensional case, for which
$\gamma_{n}\,\simeq\,1-\sqrt{2Z}$ at small $Z$, so that $\tilde S_{1}$
diverges in the limit $Z\,\rightarrow\,0$. In higher dimensions, the
modulus of $\gamma_{n}$ is always
stricty smaller than
$1$ for all $Z$, so that $\tilde S_{1}$ no more diverges in the limit
$Z\,\rightarrow\,0$.
This property is crucial since it drastically changes
the behaviour of the drift term according to the dimensionality, setting a
marked difference between the cases  $d=1$ and $d>1$ (see also section
\ref{limcont}, where the continuum limit is analyzed). This fact points out
$d=1$ as being the marginal dimension for
the drift term.  A detailed calculation yields:
\begin{equation}
    \tilde{S}_{1}(z)\, \simeq\,\I\,\left\{\frac{1}{\sqrt{2Z}}
    +\frac{1}{2} -n_{0}+ \left[\frac{n_{0}}{2} (n_{0}-1)
    - \frac{1}{8}\right] \sqrt{2Z}\right\}
    \label{S11dz0}\enspace.
\end{equation}
Similarly:
\begin{equation}
     \tilde{C_{1}}(z)\,\simeq\,
     \frac{1}{\sqrt{2Z}} + \frac{1}{2} - n_{0} +
     \left[\frac{n_{0}}{2} (n_{0}-1) - \frac{3}{8}\right]\sqrt{2Z}
     \label{C11dz0}\enspace.
\end{equation}
Using (\ref{S11dz0}) and (\ref{C11dz0}), the small-$z$ expansions of
$\tilde{\langle x_{1} \rangle} (z)$ and $\tilde{\langle x_{1}^2 \rangle} (z)$
can now be written out. Performing then the Laplace inversion yields the
following asymptotic expansions:
\begin{equation}
     \langle x_{1} \rangle (t) \,\sim\, 2 a
     \sqrt{\frac{Wt}{\pi}} + \frac{a}{2}
     + \frac{(2x_{0}-a)^{2}}{8a\sqrt{\pi W t}}+{\cal O}(t^{-1})\enspace,
     \label{x1d1as}
\end{equation}
\begin{equation}
     \langle x_{1}^2 \rangle (t) \,\sim\,  2a^2W t +
     2 a^2  \sqrt{\frac{Wt}{\pi}} + (x_{0}^{2} - ax_{0} + \frac{a^{2}}{2})
     +  \frac{1}{8}(4x_{0}^{2}-4ax_{0}-3a^2)
     \frac{1}{\sqrt{\pi Wt}}+{\cal O}(t^{-1})\enspace.
     \label{x2d1as}
\end{equation}
Note that the initial condition $x_{0}$ does not appear in the two
first-dominant terms. The mean square deviation for the coordinate
$x_{1}$ results:
\begin{equation}
     \Delta x_{1}^{2}(t)\,\equiv\,\langle x_{1}^2
     \rangle (t)-[\langle x_{1}\rangle (t)]^{2}
     \,\sim\,2\left(1-\frac{2}{\pi}\right)\,\left[a{^2}Wt +\frac{1}{2}
     \left(x_{0}-\frac{a}{2}\right)^{2}\right]+{\cal O}(t^{-1})\enspace.
     \label{deltax2d1as}
\end{equation}
The continuum limit is obtained from the lattice model by taking the
limit $W \rightarrow\infty$, $a \rightarrow 0,\,n_{0}\rightarrow+\infty$
with $a^2W=D$ and $n_{0}a=x_{0}$ finite; performing this, one finds:
\begin{equation}
     \langle x_{1} \rangle (t) \,\sim\, 2 \sqrt{\frac{Dt}{\pi}}
     + \frac{x_{0}^{2}}{2\sqrt{\pi D t}}+{\cal O}(t^{-1})
     \enspace,\hspace{20pt}
     \Delta x_{1}^2 (t)\,\,\sim\, 2\left(1-\frac{2}{\pi}\right)\,
     \left(Dt+\frac{x_{0}^{2}}{2}\right)+{\cal
O}(t^{-1})\enspace.\label{xdeltax2d1ascont}
\end{equation}
Note that the continuous limit strongly modifies the asymptotic
expansions: in this limit, the first sub-dominant term of the lattice
model drops out and the first relative corrections to the asymptotic
leading term are ${\cal O}(t^{-1})$
instead of ${\cal O}(t^{-1/2})$ in the lattice version. On another point of
view, comparison with (\ref{x2d1as}) and
(\ref{x1d1as}) shows that the first lattice corrections
vanish as the first power of the lattice constant $a$.

The above expressions, obtained as the continuous limits of the lattice
model, can be compared to the well-known results directly obtained in
the continuous framework. For a particle subjected to a reflecting
barrier and starting a distance $x_{0}>0$ apart, an elementary
calculation yields the moment-generating
function in the continuous version, which reads:
\begin{equation}
     \psi(k,t)\,=\,\frac{1}{2}[\psi_{+}(k,t)+\psi_{-}(k,t)]
     \hspace{30pt}
     \psi_{\pm}(k,t)\,=\,\left[1+\Phi\left(\pm\frac{ x_{0}}{\sqrt{4Dt}}+\I
     k\sqrt{Dt}\right)\right]\,\E^{-Dk^{2}t\,\pm\,\I kx_{0}}
     \enspace,\label{psik1d}
\end{equation}
where $\Phi$ is the Error function \cite{gradry}. By successive derivations at
$k=0$, the two first moments result:
\begin{equation}
     \langle
x_{1}\rangle(t)\,=\,x_{0}\,\Phi\left(\frac{x_{0}}{\sqrt{4Dt}}\right)
     +\sqrt{\frac{4Dt}{\pi}}\,\E^{-x_{0}^{2}/(4Dt)}\hspace{30pt}
     \langle x_{1}^{2}\rangle(t)\,=\,2Dt+x_{0}^{2}
     \enspace.\label{x1x21dcont}
\end{equation}
These expressions are exact for all $t$; by an expansion at large times, they
precisely reproduce (\ref {xdeltax2d1ascont}), as it must.


\section{The case ${\displaystyle d}\,\geq\,2$}\label{twod}

The expected irrelevance of the forbidden site as dimension is
increased will now be explicitly displayed. Among other things,
it will be seen that, for $d=2$,
$\langle\vec x^{\,2}\rangle$ and $\Delta \vec{x}^2$
contain logarithmic corrections to the
``bare'' diffusive regime, which become constant in time for $d\,>\,2$.
These results allow to state that $d=2$ is indeed the
marginal dimension for the diffusion constant characterizing the dominant term
of the asymptotic behaviour. On the other hand, a permanent shift of
the coordinate persists for any $d$, even at very large times.

Let us begin with the average coordinate $\langle x_{1}\rangle$. Contrarily
to the case $d=1$, the integral $\gamma_{1}$ is always strictly smaller
than one, since
$\E^{-(Z+1)x}I_{n}(x)\,<\,1$ for all $Z$ and $x$ real.
Using results given in the Appendix B,
one eventually obtains the following
expression for the Laplace transform of the average coordinate valid
for $|z|\,\ll\,W$:
\begin{equation}
     \tilde{\langle x_{1} \rangle} (z) \simeq \frac{x_{0}}{z}\,
      \frac{4+ (2/n_{0}^{2})- (n_{0}-1)\,Z\ln
      Z}{4- n_{0}\, Z \ln Z}
      \label{x1d2}\enspace,
\end{equation}
which gives:
\begin{equation}
     \langle x_{1} \rangle (t) \sim
\left(x_{0}+\frac{a^{2}}{2x_{0}}\right)\,\left[1-\frac{1}{16Wt}\right]+\ldots
\equiv x_{0}+\delta x(t) \label{x1d2as}\enspace.
\end{equation}

\begin{figure}
\centering\epsfig{file=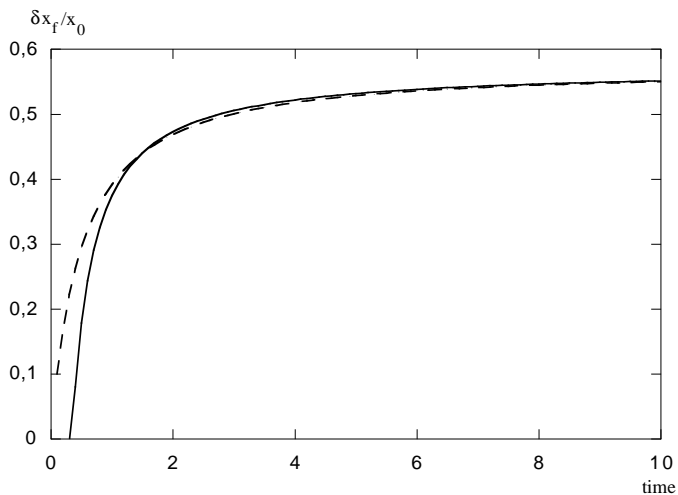,scale=0.6}
\vspace{0.3cm}
\caption{Comparison for the final shift coordinate $\delta x_{\rm f}$
between analytical (solid curve) and numerical
results (dashed line), for $n_{0}=1$ and $d=2$. The unit of time is $W^{-1}$.}
\label{fig1}
\end{figure}

\noindent The final value of the average coordinate is thus simply shifted
by $\delta
x_{\rm f}=a^{2}/(2x_{0})$, all the
more since the latter starts close to the forbidden site.
As an illustration we show on figure 1 the analytical (solid curve) result for
the final shift,
compared with the numerical calculation (dashed curve), in the case
$n_{0}=1$.
The second moment can be obtained from (\ref{fredsolC1S1}), (\ref
{fredsolCpneq1}) and (\ref{x2d}), Using the results of Appendix B:

\begin{equation}
     \tilde{\langle \vec x^{\,2} \rangle} (z) \,\simeq\, \frac{1}{z}\left[
      \frac{4a^{2}W}{z} + x_{0}^{2} +
     \frac{(Z+2)\,a_{n_{0}}(Z,\,2)}
     {1-Z\,a_{1}(Z,\,2)}\right] \enspace.\label{mom2}
\end{equation}

\begin{equation}
     \tilde{\langle \vec x^{\,2} \rangle} (z) \,\simeq\,
     \frac{1}{z}\left[\frac{4a^{2}W}{z}
     + x_{0}^{2} -\frac{a^{2}}{\pi}\ln \frac{z}{2W} \right]
     \enspace.\label{mom2as}
\end{equation}
Laplace inversion gives the dominant behaviour at large times:
\begin{equation}
     \langle \vec x^{\,2} \rangle (t) \sim 4  a^2W t +
x_{0}^{2}+\frac{a^2}{\pi} \ln 2Wt + \ldots
     \enspace.\label{mom2ast}
\end{equation}
From (\ref{x1d2as}) and (\ref{mom2ast}), one finds the mean square
deviation:
\begin{equation}
     \Delta \vec x^{\,2} (t) \sim 4  a^2W t +\frac{a^2}{\pi} \ln 2Wt
     \enspace.\label{del2ast}
\end{equation}
At this point, we see here that the diffusion
constant, defined as usually:
\begin{equation}
     D\,=\,\lim_{t\,\rightarrow\,+\infty}\,\frac{1}{2d\,t}\,
     \Delta \vec x^{\,2} (t)
     \enspace,\label{defD}
\end{equation}
is now unchanged by the
impurity but a logarithmic correction occurs in the
sub-dominant term, which can be viewed as the reminder of the change of
$D$ when $d=1$: the diffusion constant is unchanged but the first
correction diverges infinitely slowly at infinite
times. This is clearly a marginal correction: at times large enough,
the two-dimensional diffusive regime is the same with or without impurity.

In the continuous limit,
the expressions (\ref{x1d2as}) and (\ref{del2ast}) give:
\begin{equation}
     \langle x_{1} \rangle (t)\,\sim\,
     x_{0}+{\cal O}\left(\frac{a^2}{x_{0}}\right)\enspace,\hspace{30pt}
     \Delta \vec x^{\,2} (t) \,\sim\, 4 D t  +{\cal O} \left(a^2\,
     \ln \frac{2a^{2}}{Dt}\right)\hspace{30pt}
     \enspace.
\end{equation}

Let us now turn to the case $d > 2$, for which all the integrals occurring
in the subdominant terms
are strictly convergent for all $Z,\, \Re Z\,\ge\,0$, whereas the dominant
terms are exactly the same as in the absence of a localized impurity
(indeed, all the integrands behave at most as $x^{-d/2}$ at infinity).
More precisely, one finds:
\begin{equation}
     \tilde{\langle x_{1} \rangle} (z) \simeq
      \frac{x_{0}}{z} \left[1 + \frac{\xi(n_{0},\,d)}{1-
      \xi(1,\,d)} \right]\enspace,\label{x1asz}
\end{equation}
and:
\begin{equation}
      \tilde{\langle \vec x^{\,2} \rangle} (z) \simeq
     \frac{1}{z}\left( \frac{2da^{2}W}{z} + A \right)\enspace,
     \label{x2asz}
\end{equation}
where $A$ is a constant and where the function $\xi (n,\,d)$ is
defined in (\ref{defxi}). It results:
\begin{equation}
      \langle x_{1} \rangle (t)\,\sim\,x_{0} \,
\left[1+\frac{\xi(n_{0},\,d)}{1-\xi(1,\,d)}\right]\enspace,
      \hspace{30pt}\Delta \vec x^{\,2}
      (t)\,\sim\,2da^{2}Wt+A'
      \enspace,\label{x12dsup2}
\end{equation}
where $A'$ is another constant.
The correction to unity in the brackets clearly represents the microscopic
exclusion phenomemon, which again gives a final shift of the particle
position, due to the presence of
the impurity. As shown in the Appendix B, (\ref{xiapp}),
the function $\xi(n_{0},\,d)$ is approximately given by:
\begin{equation}
     \xi(n,\,d) \simeq
     (\sqrt{\pi}n)^{-d}\,\Gamma(d/2)\hspace{20pt}(n\,\gg\,1)\enspace,
     \label{deltanas}
\end{equation}
where $\Gamma$ denotes the Euler Gamma function.
Equations (\ref{x1asz}) and (\ref{deltanas}) show that the final
coordinate shift has an algebraic behaviour ($\delta x\,\sim\,
{a^{d}\,x_{0}}^{-d+1}$) as a function of the distance between the
starting point and the forbidden site. On the other hand, the subdominant
term in $\Delta \vec
x^{\,2}$ turns out to be now
a constant in time, instead of diverging logarithmically as
is the case for $d=2$.

\section{The space continuous limit}\label{limcont}
\label{tcl}

Here and there in the above, we gave the continuum limits of the first
moments, in order to emphasize the corrections in the
asymptotic regimes which are due to the lattice.
In this section, we explicitly show that the $d>1$-random walk in
continuum space is, as a whole, unaffected at all times by the local
impurity.
The central quantities  $\tilde{S_{1}}$ and $\tilde{C_{p}}$ only depend on the
reduced variable
$Z = za^2/(2D)$, pointing out that the continuum limit embodies the limit
$Z \rightarrow 0$. Let us first consider the behaviour of
$\tilde{S_{1}}$; using the
results of the Appendix B, we find at the lowest order in $Z$ :
\begin{equation}
     \tilde{S_{1}}(Z)\,\simeq\,\I \,\frac{\pi^{-d/2}
     \Gamma(d/2)n_{0}^{-(d-1)}}
     {\left[1-\pi^{-\frac{d}{2}}
     \Gamma\left(d/2\right)\right] -F_{d}(Z)}\label{Scont}\enspace,
\end{equation}
where:
\begin{equation}
         F_{1}(Z)\,\simeq\,- \sqrt{2Z}\enspace,\hspace{20pt}
         F_{2}(Z)\,\simeq\,\frac{1}{2 \pi} Z\log{Z}\enspace,\hspace{20pt}
         F_{d > 2}(Z)\,\simeq\,{\rm Cte} \times Z \enspace.
\end{equation}
Here we clearly see that for the drift term, the marginal dimension is
$d=1$: the term $\tilde{S_{1}}$ is divergent at low
$Z$ only for $d=1$ and remains finite in this limit
for $d>1$. Since the drift only depends on
 $\tilde{S_{1}}$, we can conclude that,
for $d=1$ , $\langle x_1 \rangle (t) - x_{0} =  2 \sqrt{Dt/\pi}$,
whereas for higher dimension, $\langle x_1 \rangle - x_{0}$ vanish as $a^d$
in the continuum limit.

Generally speaking, it is easy to obtain the continuous limit of the
characteristic function (\ref{solcharac}). A few algebra yields :
\begin{equation}
     \tilde{\psi}(k,z)=\frac{1}{z+Dk^2}(\psi_{0}+\I k \sqrt{\frac{D}{z}} +
{\cal O}(a))
      \hspace{30pt}(d\,=\,1)\enspace,
\end{equation}
\begin{equation}
     \tilde{\psi}(\vec{k},z)=\frac{1}{z+Dk^2}(\psi_{0}+ {\cal O}(a^d))
      \hspace{30pt}(d\,> \,1)\enspace.
\end{equation}
Thus, when the continuum limit is performed,
all the additional terms arising from
the excluded site strictly vanish except for $d\,=\,1$. As a whole,
it turns out that $d=1$ is
the marginal dimension of the problem in its space continuous version.

As a final remark, let us note that the present work indeed solves the
problem of two particles with a hard-core mutual repulsion (exclusion
process), each of them having the same diffusion constant $D$. By
separating the free diffusion of the center-of-mass -- which
undergoes a normal diffusion with a constant $D/2$ --, one is left
with a reduced particle with a constant $2D$ and subjected to a
static excluded point, which is the problem fully solved above.


\section{conclusions}

We have studied the effect of a forbidden site on the
random walk on a hypercubic lattice according to the dimensionality
$d$ of the latter. For the mean square
dispersion of the coordinate, the diffusion constant $D$ defining the
asymptotical dynamics is altered only when $d=1$; for $d=2$, logarithmic
corrections $\sim \ln t$ occur, which become constant in time when $d>2$,
whereas $D$ is unchanged. On the contrary, the
average coordinate is sensitive to the impurity in all cases. For
$d=1$, it goes to infinity at large times ($\langle x\rangle(t)\,\sim
t^{1/2}$); for $d\,\ge\,2$, its limiting value is always finite and
displays a shift $\delta x_{\rm f}\propto a^{d}x_{0}^{-d+1}$
as compared to its initial value $x_{0}$.

As expected, these effects are found to be less important in the continuous-
space version of the problem. By considering the generating function
of the moments, it was shown that the impurity is totally
irrelevant except for $d=1$: as far as $d\,>\,1$, the
continuous-space generating function is exactly the same as for the
ordinary random walk (the above-mentionned corrections for
$d\ge 2$ are thus lattice effects). This allows to state that
$d=1$ is the marginal dimension for the continuous-space problem.

All these results were obtained by assuming a contact interaction
between the walker and the impurity. Obviously enough, it can be
anticipated that a long-range interaction strongly alters the present
conclusions. Up to our knowledge, the interplay between the
dimensionality and the interaction range is, presently, an open
question that clearly deserves further study.\\\\


We are indebted to Julien Vidal for his most valuable remarks and comments
on the
manuscript.

\appendix

\section{Solution of the Fredholm equation}\label{AppendixA}

Equation (\ref{masterlap}) is a homogeneous separable Fredholm system
and can be solved by quadratures. We set:
\begin {equation}
     G(\vec\phi,z)\,=\,\frac{1}{z + 2W\sum_{p=1}^{d}(1- \cos
     \phi_{p})}\enspace,
\label{green}
\end{equation}
and introduce the auxiliary dimensionless quantities:
\begin{equation}
     \tilde{C_{p}}(z)\, =\,2W\,\int\, \D^d
     \phi\,\cos  \phi_{p}\,\tilde{\psi}(\vec\phi,z)
     \enspace,\hspace{20pt}
     \tilde{S_{1}}(z)\,=\,2W\, \int\, \D^d
     \phi\,\sin  \phi_{1}\,\tilde{\psi}(\vec\phi,z)
      \enspace.
     \label {SCP}
\end {equation}
Due to the chosen initial condition, the $(d-1)$ functions ${\tilde C}_{p}$,
$p=2,\,3,\,\ldots,\,d$, are all equal and
are simply denoted by ${\tilde C}$ in the following. Using the
standard procedure for solving such a separable Fredholm equation,
$\tilde{C_{p}}$ and $\tilde{S_{1}}$ are seen to be given by an
inhomogeneous system, which can be readily written and solved. Then, using
the identity $u^{-1}=\int_{0}^{+\infty} \D x
\,\E^{-ux},\,\,\Re u\,>\,0$, it turns out to be that
all these quantities can be expressed in terms of
integrals involving the
Bessel functions of the second kind $I_{n}$. More precisely:
\begin {equation}
     a_{n}(Z,\,d)\,=\,
     \int_{0}^{+\infty} \D x \, \E^{-(Z+d)x} \,
\BI_{n}(x) \, \BI_{0}^{d-1}(x)\,\hspace{20pt}
     b_{nn'}(Z,\,d)\,=\,
     \int_{0}^{+\infty} \D x \, \E^{-(Z+d)x} \,
\BI_{n}(x) \, \BI_{n'}(x) \, \BI_{0}^{d-1}(x)\,
\enspace,\label{alpbetgam0}
\end {equation}
A somewhat lengthy but straightforward calculation gives:
\begin{equation}
\tilde{C_{1}}\,=\,\frac{1}{D}\,[1+a_{1}-\beta_{1}-(d-2)(\alpha_{11}-a_{1})]
     \beta_{n_{0}}+(d-1)(\alpha_{11}-a_{1})\alpha_{n_{0}1}]\enspace,
\hspace{30pt}
     \tilde{S_{1}}\,=\,\frac{\I\gamma_{n_{0}}}{1-\gamma_{1}}\enspace,
     \label{fredsolC1S1}
\end{equation}
\begin{equation}
     \tilde{C}_{p\neq 1}\,\equiv\,\tilde{C}\,=\,
     \frac{1}{D}\,[(1+a_{1}-\beta_{1})\alpha_{n_{0}1}
     +(\alpha_{11}-a_{1})\beta_{n_{0}}]\enspace.
\label{fredsolCpneq1}
\end{equation}
The various quantities appearing in these expressions are defined as
follows:
\begin{equation}
     D\,=\,[1+a_{1}-\beta_{1}-(d-2)(\alpha_{11}-a_{1})](1+a_{1}
     -\beta_{1})-(d-1)(\alpha_{11}-a_{1})^{2}]
     \enspace,\label{determ}
\end{equation}
\begin {equation}
     \beta_{n}\,=\,\frac{1}{2}\,[a_{n-1}+a_{n+1}] \enspace,
     \hspace{25pt}\gamma_{n}\,=\,\frac{1}{2}\,[a_{n-1}-a_{n+1}]\enspace,
     \hspace{25pt}\alpha_{nn'}\,=\,b_{nn'}(Z,\,d-1)\enspace.
     \label{fredbessela0b0}
\end{equation}
With these definitions, the solution of the central equation
(\ref{masterlap}) writes:
\begin {equation}
     \tilde {\psi} (\vec{\phi},z) \,=\,
     G(\vec\phi,z)\,\left[\psi_{0}+
     \,\tilde{C}(z)\,\sum_{p=2}^{d}(\cos\phi_{p}- 1)\, +
     \tilde{C_{1}}(z)\, (\cos\phi_{1}- 1) +\tilde{S_{1}}(z)\,\sin
\phi_{1}\,\right]
     \enspace,
     \label{solcharac}
\end {equation}
and is the starting point of the analysis undertaken in the present
work.


\section{Asymptotical behaviour of many-Bessel functions
integrals}
\label{appendix B}

In this paper we repeatedly have to estimate integrals of the kind:

\begin{equation}
     I(Z,d,\vec{n},m) = \int_{0}^{\infty} \D x \,
     \E^{-(Z+d)x} I_{n_{1}}(x) \ldots I_{n_{d}}(x) \,\,\, x^m
     \enspace,\label{defintB}
\end{equation}

in the limit $Z\,\ll\,1$, where $I_{n}(x)$ denotes a modified Bessel function
and $\vec{n}=\{n_{1},\ldots,n_{d}\}$.
When $Z =0$, no singularity arises from the lower bound
provided that $ n_{1}+ \ldots + n_{d} + m > -1$
( this is always the case in this paper). On the other side
($x \rightarrow \infty$), the integral is finite for $m < (d/2) -1$.
Our goal is to obtain the asymptotical behaviour of $I(Z,d,\vec{n},m)$
for large $n_{i}$ (but numerical checks show that some of the following
analytical results still hold true even for $n^2=\Sigma n_{i}^{2}=1$).

In particular, when $I(Z,d,\vec{n},m)$  is divergent, the knowledge of
the nature of the
divergence for $Z \sim 0$ is required. Let us first introduce the
generalized integrals :
\begin{equation}
      I_{g} (Z,\,d,\,\vec{n},\,m, \,\omega) =  \int_{0}^{\infty} \D x \,\,\,
      \E^{-(Z+d)x} \E^{-\omega/x}\,
      I_{n_{1}}(x) \ldots I_{n_{d}}(x) \, x^m\enspace.
\end{equation}
which share the same diverging properties than $I$ when $Z\,\ll\,1$ and
allow to find
the latter by one or other of the following procedures:
\begin{equation}
      I(Z,d,\vec{n},\,m) = \lim_{\omega \rightarrow 0}
      I_{g} (Z,\,d,\,\vec{n},\,m,\, \omega)\enspace.
\end{equation}
\begin{equation}
      I_{g} (Z,\,d,\,\vec{n},\,m-1,\,\omega) = - \frac{\partial}{\partial
\omega}
      I_{g} (Z,\,d,\,\vec{n},\,m,\,\,\omega)
      \label{ascend} \enspace,
\end{equation}
\begin{equation}
      I_{g} (Z,\,d,\,\vec{n},\,m+1,\,\omega) = - \frac{\partial}{\partial Z}
      I_{g} (Z,\,d,\,\vec{n},\,m,\,\omega)
      \label{descend} \enspace.
\end{equation}

Let us first examine the case $d$ even ($d=2p$). The first
non-logarithmically diverging integral is $I_{g} (Z,2p,\vec{n},p,\omega)$.
The divergence near $Z\,=\,0$ arises from large values of $x$. In order
to get for purpose a convenient
asymptotical form of Bessel functions,
we first use an integral representation of the latter, and approximate
it by using a
stationary-phase argument when both $n$ and $x$ are much greater than unity :
\begin{equation}
      I_{n}(x) = \frac{1}{2 \pi} \int_{-\pi}^{+\pi} \D \theta  \,
      \E^{x \cos\theta + \I n \theta}
      \simeq \frac{1}{2 \pi} \int_{-\infty}^{\infty}\,\D \theta  \,
       \E^{x(1-\frac{1}{2} \theta^2)+\I n \theta}
       = \frac{1}{\sqrt{2 \pi x}} \,\E^{\,x-\frac{n^2}{2x}}
       \hspace{20pt}(x,\,n\,\gg\,1)\label{Bessstph}\enspace.
\end{equation}
This approximation being done, all the integrals $I_{g}$ and $I$ only depend
on the length of the vector $\vec n$, simply denoted by $n$ in the
following. Using the approximate expression (\ref{Bessstph}), we
obtain with help of \cite{gradry} :
\begin{equation}
      I_{g} (Z,\,2p,\,\vec{n},\,p,\,\omega) \simeq (2 \pi)^{-p}
      \int_{0}^{\infty} \D x \,\,\, \E^{-Z x} \E^{-\frac{\omega +
      n^2/2}{x}} = (2 \pi)^{-p}
      \sqrt{\frac{4(\omega + n^2/2)}{Z}}\,
      K_{1} \left( \sqrt{4(\omega + n^2/2) Z}
      \right)\enspace,\label{Igaseven}
\end{equation}
$K_{1}$ denoting the Bessel function of the second kind.

For $d$ odd ($d=2p+1$) we choose $I_{g}(Z,2p+1,\vec{n},p,\omega)$
as a generator; by the same way as above we find:
\begin{equation}
      I_{g} (Z,\,2p+1,\,\vec{n},\,p,\,\omega) \simeq (2 \pi)^{-(p+\frac{1}{2})}
      \int_{0}^{\infty} \D x \,\,\, \E^{-Z x^2}
      \E^{-\frac{\omega+n^2/2}{x^2}} = (2 \pi)^{-(p-\frac{1}{2})}
      \sqrt{\frac{\pi}{Z}} \,\,\E^{- 2 \sqrt{ Z (\omega +
      n^2/2})} \enspace.\label{Igasodd}
\end{equation}
Expressions (\ref{Igaseven}) and (\ref{Igasodd}) are correct in the
limit $Z\,\ll\,1$ and $n^{\,2}\,\gg\,1$.
Using then (\ref{ascend}) and (\ref{descend}), we are now able to calculate
the asymptotical behaviour of the whole family of $I$ integrals.

As an example, let us consider $I(Z,\,2,\,\{0,n\},\,-1)$,
which occurs in the calculation of the average coordinate for $d=2$.
Explicitly, one has:
\begin{equation}
      I(Z,\,2,\,\{0,n\},\,-1)\,=\, \int_{0}^{\infty} \D x \,
x^{-1}\,\E^{-(Z+2) x}
     \, I_{0}(x) \,I_{n}(x)  \enspace,
\end{equation}
and the method sketched above yields:
\begin{equation}
      I(Z,\,2,\,\{0,n\},\,-1)\,\simeq\,\frac{1}{\pi n^{2}} +
      \frac{1}{2 \pi} Z \ln Z
      \hspace{20pt}(Z\,\ll\,1 \,\, , \,\, n \, \gg \, 1)\enspace.
      \label{IZ2-1)}
\end{equation}
We numerically checked the value of the $Z$-independent constant; it
turns out that for $n\,\ge\,5$, $1/(\pi n^{2})$ gives the correct value
with a relative error smaller than $\sim 10^{-3}$.
On the other hand,
we found no simple way to numerically check the validity of the
first-order term in the expansion (\ref {IZ2-1)}), but one
can analytically prove that it is correct for $I(Z,2,\{0,1\},-1)$.
Indeed, the integral  $I(Z,2,\{0,1\},1)$ is computed in
\cite{gradry}:
\begin{equation}
      I(Z,\,2,\,\{0,1\},\,1) = \frac{1}{2 \pi} \left[
      \frac{Z'}{{Z'}^{2}-1} {\bf E}(\frac{1}{Z'})-\frac{1}{Z'}{\bf K}(
      \frac{1}{Z'}) \right] \enspace,
\end{equation}
where ${\bf K}$ and ${\bf E}$ are the complete elliptic
functions of the first and the second kind respectively, and where $Z' = 1 +
Z/2$. After a few algebra, we find :
\begin{equation}
        \lim_{Z\,\rightarrow\,0} Z\,I(Z,\,2,\,\{0,1\},\,1) =
        \frac{1}{2 \pi} \label{LimIZ0}\enspace.
\end{equation}
Setting:
\begin{equation}
       \xi(\vec n,\,d)\,=\,I(0,\,d,\,\vec n,\,-1)
       \label{defxi}\enspace,
\end{equation}
(\ref {LimIZ0}) entails:
\begin{equation}
       \lim_{Z\,\rightarrow\,0} I(Z,\,2,\,\{0,1\},\,-1) =
       \xi(\{0,1\},\, 2)+ \frac{1}{2 \pi} Z \ln Z + C_{1} Z + \ldots
       \enspace,
\end{equation}
which for $n=1$ is in agreement with (\ref {IZ2-1)}).

As an application, let us find the asymptotical behaviour of the
function $\xi(n,\,d)$ defined by (\ref{defxi}), which is strongly related
to the coordinate shift in any dimension. For $d=2p$, we find
(remember that $n\,=\,|\vec n|$):
\begin{equation}
      \xi(n,2p)\,=\, {(-1)}^{p+1} \left.
      {\frac{{\partial}^{p+1}}{\partial {\omega}^{p+1}}
      I_{g}(Z,\,\vec{n},\,2p,\,p,\,\omega)}
      \right|_{Z=\omega=0}\,\simeq\,{\pi}^{-p} (p-1)! \,n^{-2p} \enspace,
\end{equation}
whereas for $d=2p+1$, one has:
\begin{equation}
       \xi(n,2p+1) ={(-1)}^{p+1} \left. {\frac{{\partial}^{p+1}}
       {\partial {\omega}^{p+1}}
       I_{g}(Z,\,\vec{n},\,2p+1,\,p,\,\omega)}
       \right|_{Z=\omega=0}\,\simeq\,{\pi}^{-p}
       \frac{1.3.5 \ldots(2p-1)}{2^{p}} \,n^{-(2p+1)}
        \enspace. \end{equation}
These two definitions turn in fact into only one :
\begin{equation}
    \xi(n,d)\,\simeq\,{\pi}^{-\frac{d}{2}}\, \Gamma(d/2)\,
     n^{-d}
     \enspace,\label{xiapp}
\end{equation}
$\Gamma$ still denoting the Euler Gamma function.
It is worth noting that this asymptotical behaviour,
{\it a priori} only valid for $|\vec{n}\,| \gg 1$, is very rapidly
convergent and actually holds true even for rather small values of
$|\vec{n}\,|$, as shown by a few numerical checks.



\end{document}